\documentclass[
    aps,
    prl,
    reprint,
    superscriptaddress,
    longbibliography,
    amsmath,
    amssymb
]{revtex4-2}

\usepackage{graphicx}
\usepackage{dcolumn}
\usepackage{bm}
\usepackage{hyperref}
\usepackage{booktabs} 

\begin{document}

\preprint{APS/123-QED}

\title{Superselectivity and the Thermodynamics of Biomolecular Condensation}

\author{Charley Schaefer}%
\email{c.schaefer@leeds.ac.uk}
\affiliation{%
  School of Physics and Astronomy, University of Leeds, UK
}%
\affiliation{%
   Astbury Centre for Structural Molecular Biology, University of Leeds, UK
}%

\begin{abstract}
Multivalent binding drives superselective adsorption and biomolecular condensation, yet their connection remains unclear. We develop a mean-field theory for multivalent linker polymers associating to three-dimensionally distributed binding sites, validated against Monte Carlo simulations. The theory establishes how the linker-mediated contribution to the free-energy curvature relevant to condensation can be calculated from density-dependent adsorption isotherms.
We find that linker-mediated attractions develop at intermediate linker concentrations and are suppressed at higher concentrations by competition for binding sites, even though molecular binding persists.
Applied to the EPYC1--Rubisco pyrenoid, the theory predicts how sticker number and spacer length control the strength and concentration range of these interactions.

\end{abstract}

\maketitle
Multivalent binding is ubiquitous in nature and technology \cite{choi_physical_2020, Semenov1998, PhysRevLett.97.168301,Bianchi2011, Seiffert2015}, and gives rise to two related equilibrium phenomena. First, multivalent molecules can discriminate between environments with different densities of binding sites: a small increase in site density can lead to a disproportionately large increase in adsorption through combinatorial advantages, a phenomenon known as superselectivity \cite{mammen1998polyvalent, Frenkel2011, Dubacheva2014-yl, Miller21, xie2025receptor}. Second, these molecules can crosslink their binding partners, generating effective attractions that can drive liquid--liquid phase separation and gelation of (biomolecular) condensates \cite{Semenov1998,choi_physical_2020, banani_biomolecular_2017}. While the interactions underlying condensation can be quantified experimentally \cite{ausserwoger_quantifying_2025}, it remains unclear how these attractions are related to the combinatorial advantages that give rise to superselectivity, particularly when competition for binding sites suppresses multivalent binding at higher linker concentrations \cite{PhysRevLett.132.118202}. Here, we establish how the effective attractions generated by multivalent molecules can be inferred from their (density-dependent) adsorption statistics, and investigate how these attractions evolve across the crossover from superselective to competitive binding.

\begin{figure}
\includegraphics[width=8cm]{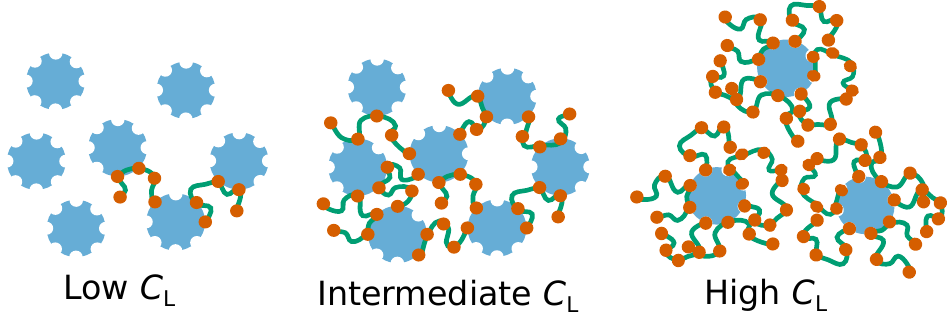}
\caption{Competition between multivalent binding and
binding-site saturation, illustrated for EPYC1--Rubisco. Each EPYC1 linker contains five stickers, while
each Rubisco molecule presents eight binding sites.
At low linker concentrations $C_\mathrm{L}$, individual
linkers can bind through multiple stickers, but few
crosslinks are formed.
At intermediate concentrations, the increasing number
of crosslinks generates appreciable effective attractions.
At high concentrations, individual binding sites become
occupied by separate linker molecules, suppressing
crosslinking. We investigate how these regimes are
thermodynamically related to adsorption selectivity.}
\label{fig:Illustration}
\end{figure}

We address this by developing a statistical-mechanical
theory of multivalent binding, using the algal pyrenoid
as a model system \cite{schaefer2026biophysicspyrenoid}, see Fig.~\ref{fig:Illustration}. The pyrenoid is a biomolecular
condensate that supercharges photosynthesis in algae
\cite{barrett_pyrenoids_2021},
in which multivalent interactions can be investigated
at both the molecular level and the level of condensate
formation \cite{payne-dwyer_predicting_2024,Kumar2026.01.27.701992}.
Its matrix consists primarily of Rubisco holoenzymes, each presenting eight binding sites, crosslinked by the biopolymer `linker' EPYC1 \cite{Mackinder2016}, which contains five binding motifs or `stickers' \cite{freeman_rosenzweig_eukaryotic_2017}. The sticker-and-spacer framework connects the combinatorics of these interactions to the conformational entropy of the intervening flexible `spacers' \cite{freeman_rosenzweig_eukaryotic_2017,GrandPre2023,payne-dwyer_predicting_2024}.

At the condensate level, phenomenological mean-field theories (e.g., Flory-Huggins theory\cite{flory1953principles}) provide a description of liquid--liquid phase separation and its dependence on molecular concentrations and effective interactions \cite{Brangwynne}. Previous mean-field models have successfully described aspects of pyrenoid size control \cite{He2026}, as well as the dependence of condensation thresholds on molecular weight and concentration \cite{schaefer2026biophysicspyrenoid}. 
However, without an explicit treatment of multivalent binding, these models do not directly relate the observed dependence on spacer length and sticker valency to the underlying molecular interactions \cite{GrandPre2023, Kumar2026.01.27.701992, schaefer2026biophysicspyrenoid}. In particular, EPYC1 variants with three stickers can induce condensation in vitro, whereas at least four stickers are required for pyrenoid formation in vivo \cite{Kumar2026.01.27.701992}. Moreover, changes in sticker number can affect the threshold for condensation more strongly than the properties of the resulting condensate \cite{Kumar2026.01.27.701992}. These observations highlight the broader challenge of connecting the molecular architecture of disordered proteins to the thermodynamics of concentrated biomolecular systems.

Here, we build on our recent statistical-mechanical model of multivalent polymer binding to target molecules \cite{payne-dwyer_predicting_2024}, and develop a mean-field theory that preserves the impact of binding combinatorics, competition for binding sites, and spacer conformational entropy. 
We validate the theory against Monte Carlo simulations and derive scaling laws for the crossover between superselective and competitive binding. 
Crucially, we show how the linker-mediated attraction can be calculated from molecular adsorption isotherms across these binding regimes.
Throughout this work, we illustrate our findings by applying the theory to the EPYC1--Rubisco system, and draw conclusions about the extent of superselectivity and competition in that system, while connecting the molecular properties to the linker-mediated contribution to the free energy.

\vspace{0.2em}
\paragraph{Molecular Equilibrium Statistics --}~We consider \(M\) multivalent linker polymers containing $S$ stickers that are bound with \(B\) stickers to \(N\) lattice sites using a binding energy \(\varepsilon\); thus \(N\geq B \geq M\). Whenever two stickers \(i\) and \(j\) of the same molecule bind at a distance \(z\), and any sticker in between is unbound, the elastic free energy of the {total} spacer with length \(L_{i,j}\) is described by the freely-jointed chain model \cite{doi1986theory},
\begin{align}
\frac{G(z;L_{i,j})}{k_\mathrm{B}T}
={}&
\frac{3}{2}\ln\left(
\frac{2\pi\gamma L_{i,j}}{3l_\mathrm{K}}
\right)
\notag\\
&+
\frac{L_{i,j}}{l_\mathrm{K}}
\left[
\frac{1}{2}\left(\frac{z}{L_{i,j}}\right)^2
-\ln\left(1-\left(\frac{z}{L_{i,j}}\right)^2\right)
\right],
\label{eq:StretchPotential}
\end{align}
with $l_\mathrm{K}$ the Kuhn length, and $\gamma$ a non-universal constant that we set to unity \cite{payne-dwyer_predicting_2024}. The exchange of molecules between the system and a reservoir with unbound molecules is captured using the chemical potential 
\begin{equation}
  \mu = \varepsilon+k_\mathrm{B}T \ln(C_\mathrm{L}/K_\mathrm{d}),\label{eq:chemical_potential}
  \end{equation}where the dissociation constant of a monovalent interaction  \(K_\mathrm{d}\propto \exp(\varepsilon/k_\mathrm{B}T)\) allows for the 
parametrisation of the model using single-sticker constructs \cite{he_structural_2020, GrandPre2023}.

Indeed, this model allows for Monte Carlo sampling using molecular adsorption/desorption,  sticker adsorption/desorption, and bondswapping \cite{payne-dwyer_predicting_2024}, which generates expectation values for the per-site coverages \(\theta_\mathrm{m}\equiv M/N\) and  \(\theta_\mathrm{s}\equiv B/N\).
For small systems, these can be analytically calculated using the  configurational partition sum,
\begin{equation}
Z_{M,B}=
\frac{N!}{(N-B)!}
\sum_{\mathbf m\in\mathcal P_{M,B}}
W_{\mathbf m}
\prod_{b=1}^{S}
\frac{\binom Sb^{m_b}}{m_b!},
\label{eq:ZMB}
\end{equation}
where $\mathcal P_{M,B}$ denotes the allowed integer partitions $\mathbf m=[m_1,\ldots,m_S]$, with $m_b$ counting the number of molecules bound through $b$ stickers. The conformational weight
\begin{equation}
W_{\mathbf m}=
\psi_{\mathbf m}
\left\langle
\exp\left(-\frac{G_\mathrm{elas}}{k_\mathrm{B}T}\right)
\right\rangle_{\mathbf m}
\label{eq:ConformationalWeight}
\end{equation}
is determined numerically through Monte Carlo sampling, and
accounts for the restrictions on accessible configurations and the elastic free energy of the spacers. The factor $N!/(N-B)!$ ensures that each binding site accommodates at most one sticker.

\vspace{0.2em}

\begin{figure}[ht!]
\includegraphics[width=9cm]{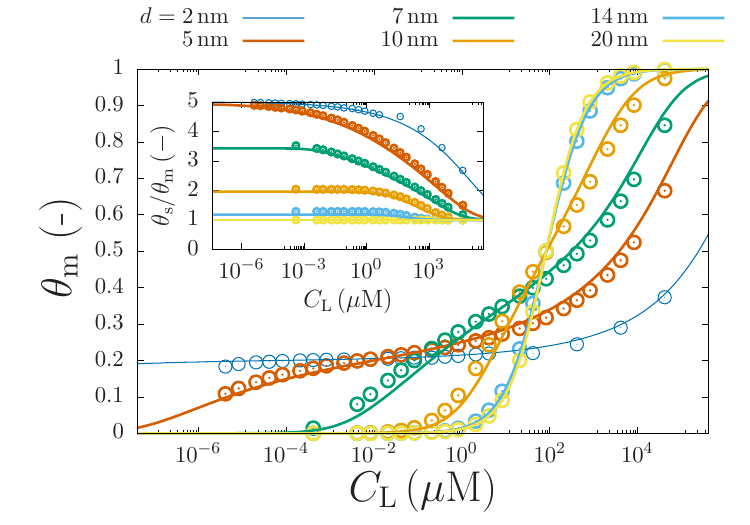}
\caption{Adsorption isotherms of molecular coverage of a condensate (main panel) and the bound stickers per bound linker (inset). The symbols represent the simulations while the solid curves represent the mean-field theory in Eqs.~(\ref{eq:StickerAdsorption}-\ref{eq:MoleculeAdsorption}).
}\label{fig:adsorption}
\end{figure}

\paragraph{Simulations --}~
Unless stated differently, throughout this work  we use  the physical parameters  that describe the EPYC1:Rubisco system \cite{payne-dwyer_predicting_2024}: The linker molecule has  $S=5$ stickers, spacer contour length $L=18\,\mathrm{nm}$, Kuhn length $l_\mathrm{K}=0.88\,\mathrm{nm}$,  sticker binding energy $\varepsilon=-11.8\,k_\mathrm{B}T$, and single-sticker dissociation constant \(K_\mathrm{d}=414\,\mathrm{\mu M}\). 
While the sites on Rubisco are at a distance 7.7 nm \cite{he_structural_2020, payne-dwyer_predicting_2024}, and the intersite distance between nearest Rubisco neighbours is 4 nm \cite{he_structural_2020}, we here neglect these hetereogeneities, and use \(d=7\,\mathrm{nm}\) as lattice spacing to reproduce the correct macroscopic concentration of Rubisco in the pyrenoid (i.e., the site density is \(\rho\approx 3\times 10^{-3}\,\mathrm{nm}^{-3}\)) \cite{mackinder_spatial_2017}.
The simulation results for various site densities and unbound-linker concentrations are shown as symbols in  Fig~\ref{fig:adsorption}, while the solid curves represent the mean-field theory that is derived in the following.

\vspace{0.2em}
\paragraph{Mean-field theory --}~
We first treat the conformations of different bound molecules as independent, such that
\begin{equation}
W_{\mathbf m}\simeq
\prod_{b=1}^{S}w_b^{m_b},
\label{eq:LinkerFactorisation}
\end{equation}
where $w_b$ is the conformational weight of a molecule bound through $b$ stickers. Second, we treat the constrained spacers of an individual molecule independently, neglecting correlations between their spatial configurations. Third, we approximate the binding sites as uniformly distributed within a volume $V$, with number density $\rho=N/V$ and characteristic separation $d=\rho^{-1/3}$. This approximation will lead to accurate predictions for the calculation of the coverages (see Fig.~\ref{fig:adsorption})  and grand potential (Supplementary Information [SI]) that we derive below, while the effective attractions related to condensation (also derived below) are somewhat more sensitive to the local site distribution, see SI.

Under these approximations, the spatially averaged Boltzmann factor of an individual constrained spacer is described by the conformational volume
\begin{equation}
v_L=
4\pi\int_d^L
\mathrm{d}z\,z^2
\exp\left(-\frac{G(z;L)}{k_\mathrm{B}T}\right).
\label{eq:ConformationalVolume}
\end{equation}
The lower integration boundary provides an approximate description of the finite separation between accessible binding sites. For \(d\geq L\), we have \(\upsilon_\mathrm{L}=0\).


The conformational weight of a multivalent molecule follows by summing over the possible choices of bound stickers.  In the SI, we show this leads to the free energy per binding site\cite{chaikin1995principles}
\begin{align}
\frac{f}{k_\mathrm{B}T}
={}&
(1-\theta_\mathrm{s})
\ln(1-\theta_\mathrm{s})
+\theta_\mathrm{s}-\theta_\mathrm{m}
\notag\\
&+
\sum_{b=1}^{S}x_b
\ln\left(
\frac{x_b}{ q_b}
\right)
+\frac{\varepsilon}{k_\mathrm{B}T}
\theta_\mathrm{s},
\label{eq:FreeEnergy}
\end{align}
where \(x_b=m_b/N\) measures the adsorption of linker molecules using \(b\) stickers per molecule, and 
\[
q_b\equiv N^{b-1}\binom{S}{b} w_b =  \sum_{i=b}^ST_{i,b}.
\]
In this expression, \(T_{i,b}\) is calculated recursively using \(T_{i,1}=1\) and
\(
T_{i,b}=\sum_{j=1}^{i-1}T_{j,{b-1}}\,\rho \,\upsilon_{ L_{j, i}}
\). 

To find the equilibrium coverages, we minimise the grand potential \(w=f-\mu\theta_\mathrm{m}\), where the chemical potential of the linker molecules is given in Eq.~\ref{eq:chemical_potential}.
Hence, the minimisation of \(w\) yields
\begin{equation}
x_b=
\frac{C_\mathrm{L}}{K_\mathrm{d}} q_b
(1-\theta_\mathrm{s})^b
\exp\left(
-\frac{(b-1)\varepsilon}{k_\mathrm{B}T}
\right).
\label{eq:xb}
\end{equation}
The factor $(1-\theta_\mathrm{s})^b$ accounts for the depletion of available binding sites as the coverage increases.

Substituting the equilibrium distribution, Eq.~(\ref{eq:xb}), into Eq.~(\ref{eq:FreeEnergy}), and using the definitions of \(\theta_\mathrm{m}\) and \(\theta_\mathrm{s}\), we obtain
\[
\begin{aligned}
\sum_{b=1}^{S}x_b\ln\left(\frac{x_b}{q_b}\right)
={}&\theta_\mathrm{m}\ln \frac{C_\mathrm{L}}{K_\mathrm{d}}
+\theta_\mathrm{s}\ln(1-\theta_\mathrm{s})
-\frac{(\theta_\mathrm{s}-\theta_\mathrm{m})\varepsilon}{k_\mathrm{B}T}
.
\end{aligned}
\]
The equilibrium free energy per binding site therefore simplifies to
\begin{equation}
\frac{f}{k_\mathrm{B}T}
=
\ln(1-\theta_\mathrm{s})
+\theta_\mathrm{s}-\theta_\mathrm{m}
+\left(
\ln \frac{C_\mathrm{L}}{K_\mathrm{d}}
+\frac{\varepsilon}{k_\mathrm{B}T}
\right)\theta_\mathrm{m}.
\label{eq:EquilibriumFreeEnergy}
\end{equation}
Using the definitions of the molecular and sticker coverages, \(\theta_\mathrm{s}=\sum_bbx_b\) and  \(\theta_\mathrm{m}=\sum_bx_b\) give
\begin{align}
\theta_\mathrm{s}
&=\frac{C_\mathrm{L}}{K_\mathrm{d}}\sum_{b=1}^{S}
bq_b(1-\theta_\mathrm{s})^b
e^{-(b-1)\varepsilon/k_\mathrm{B}T},
\label{eq:StickerAdsorption}\\
\theta_\mathrm{m}
&=\frac{C_\mathrm{L}}{K_\mathrm{d}}\sum_{b=1}^{S}
q_b(1-\theta_\mathrm{s})^b
e^{-(b-1)\varepsilon/k_\mathrm{B}T}.
\label{eq:MoleculeAdsorption}
\end{align}
These equations are shown as solid curves in Fig.~\ref{fig:adsorption} and agree with the simulations (symbols). 

The adsorption isotherms in Fig.~\ref{fig:adsorption}
illustrate the competition between multivalent binding
and binding-site saturation.
At small site separations, linkers initially bind
through multiple stickers, such that a large fraction
of the binding sites can be occupied by relatively
few molecules.
Increasing the linker concentration favours the
adsorption of additional molecules at the expense
of the number of bound stickers per molecule (inset).
This competition gives rise to the distinct adsorption
regimes discussed below.

\vspace{0.2em}
\paragraph{Superselective and Competitive Binding--}~
The variations in adsorption mechanism are quantitied by the selectivity factor
\begin{equation}
\alpha_\mathrm{m}=
\left.
\frac{\partial\ln\theta_\mathrm{m}}
{\partial\ln\rho}
\right|_{c_\mathrm{L}}.
\label{eq:Selectivity}
\end{equation}
Indeed, binding is superselective if \(\alpha_\mathrm{m}>1\).
Values smaller than \(-1\) would point at superselective desorption.
However, here we find values \(-1<\alpha <0\) and refer to this regime as competitive binding.

\begin{figure}[ht!]
\includegraphics[width=8.5cm]{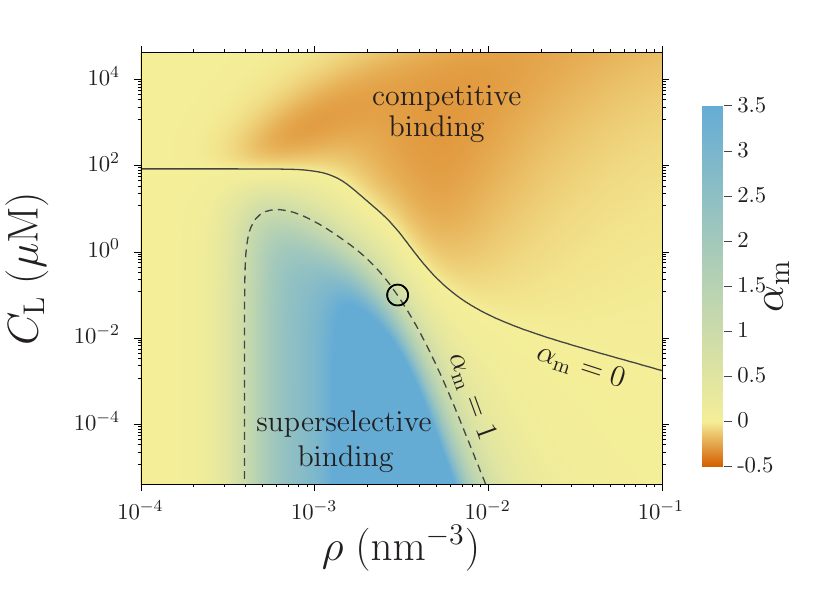}
\caption{Contour plot of the selectivity factor for various site densities and concentration of unbound linker. The circle represents our estimated site density and unbound linker concentration in the native EPYC1:Rubisco pyrenoid.
}\label{fig:SuperselectivityDiagram}
\end{figure}

A representative diagram that shows these regimes was calculated for the EPYC1:Rubisco pyrenoid system, and is displayed in  Fig.~\ref{fig:SuperselectivityDiagram}.
For reference, the experimental site density in pyrenoids is approximately \(3\times 10^{-3}\,\mathrm{nm}^{-3}\) \cite{mackinder_spatial_2017}, and the total concentration of linker molecules is about \(800\,\mathrm{\mu M}\); our calculations estimate the concentration of unbound linkers is of the order of \(C_\mathrm{L} = C_\mathrm{L}^{\mathrm{tot}}-\rho\,\theta_\mathrm{m}\approx0.1\,\mathrm{\mu M}\), i.e., in the cross-over regime between superselective- and competive binding.

To describe the crossovers between the different regimes in the diagram, we will use dimensionless units (thus, we use \(c_\mathrm{L}=C_\mathrm{L}/K_\mathrm{d}\)), and first focus on the non-universal crossover at low concentrations from the monovalent-binding regime to superselective binding, i.e, crossing the vertical portion of the dashed curve in Fig.~\ref{fig:SuperselectivityDiagram}. Indeed, the site density at which this happens is obtained by numerically solving
\[
1=\alpha_\mathrm{m}\simeq\left.\frac{\partial \ln\left(\sum_b q_b\exp(-b\varepsilon/k_\mathrm{B}T)\right)}{\partial \ln \rho} \right|_{c_\mathrm{L}},
\]
While this crossover is non-universal, all other crossovers are described by asymptotic scaling relations. For low site densities \(\rho\), the intersite distance exceeds the total polymer contour length \(d>(S-1)L\), and only a single sticker of each linker molecule can bind. In this limit, we get the usual Langmuir isotherm
\(\theta_\mathrm{m}\approx {Sc_\mathrm{L}}/({1+Sc_\mathrm{L}})\)
and the onset of competitive binding (horizontal portion of the solid curve  Fig.~\ref{fig:SuperselectivityDiagram}) is  given by
\[
c_\mathrm{L}^\ast=\frac{1}{S},\text{ for }\alpha_\mathrm{m}=0 \text{ and }\rho\ll \upsilon_\mathrm{L}^{-1}\exp(\varepsilon/k_\mathrm{B}T).
\]

For sufficiently high site densities and long, flexible spacers, the conformational volume becomes approximately independent of the contour length separating two bound stickers, \(v_{L_{i,j}}\simeq v_L\simeq l_\mathrm{K}^3/\gamma^{3/2}\). 
For finite site densities and spacer lengths, we take
$v_L$ to be the conformational volume of a spacer
connecting two adjacent stickers, evaluated using
Eq.~(\ref{eq:ConformationalVolume}).
In this case, we obtain
\begin{equation}
  q_b\simeq \binom Sb\left({\rho v_L}\right)^{b-1}.
  \label{eq:qbSimple}
\end{equation}
Introducing the dimensionless parameter
\begin{equation}
  K=\rho v_L
  \exp\left(-\frac{\varepsilon}{k_\mathrm{B}T}\right),
  \label{eq:MultivalencyParameter}
\end{equation}
which quantifies the statistical weight associated with
binding an additional sticker, we may evaluate the sums in
Eqs.~(\ref{eq:StickerAdsorption}) and
(\ref{eq:MoleculeAdsorption}) analytically to obtain
\begin{align}
  \theta_\mathrm{s}&=
  Sc_\mathrm{L}(1-\theta_\mathrm{s})
  \left[1+K(1-\theta_\mathrm{s})\right]^{S-1},
  \label{eq:StickerAdsorptionSimple}\\
  \theta_\mathrm{m}&=
  \frac{c_\mathrm{L}}{K}
  \left\{
  \left[1+K(1-\theta_\mathrm{s})\right]^S-1
  \right\}.
  \label{eq:MoleculeAdsorptionSimple}
\end{align}

From these equations we find 
\begin{equation}
c_\mathrm{L}^\ast \simeq \frac{(S-1)^{S/2}}{SK^{(S-2)/2}} , \text{ if }\alpha_\mathrm{m}=0, K\gg 1.
\end{equation}
and
\begin{equation}
c_L^*\simeq
\frac{(S-2)[2(S-1)]^{S-1}}{S^{S+1}K^{S-1}}
,\text{ if }\alpha_\mathrm{m}=1,K\gg1,
\end{equation}
which describe the solid and dashed curves, respectively, in the high site-density regime of Fig.~\ref{fig:SuperselectivityDiagram}.

\vspace{0.2em}
\paragraph{Linker-Mediated Attractions --}~
Thus, in general titration curves as illustrated in Fig.~\ref{fig:adsorption} may include both superselective- and competitive binding regimes.
Here, we connect effective attractions in condensates to the titration curves, and, therefore, also to the superselective and competitive mechanics of binding.
Thus, we consider a fixed chemical potential of the linker molecules,
such that we may write the corresponding semi-grand free-energy density as
\begin{equation}
    f_\mathrm{eff}=f_0+\rho\omega.
    \label{eq:EffectiveBinaryFreeEnergy}
\end{equation}
Here,  $f_0$ describes the reference mixture in the
absence of linker-mediated association, and $\omega$
is the equilibrium grand potential per binding site.
The key idea is that the density fluctuations underpinning condensation
are driven by a negative curvature of the association free
energy. Therefore, we quantify the effective attraction by
\begin{equation}
    A_\mathrm{eff}=
    -\frac{1}{2}
    \left.
    \frac{\partial^2}{\partial\rho^2}
    \left(\frac{\rho\omega}{k_\mathrm{B}T}\right)
    \right|_{c_\mathrm{L}}.
    \label{eq:EffectiveAttraction}
\end{equation}
Using the thermodynamic identity
$\partial(\beta\omega)/\partial\ln c_\mathrm{L}
=-\theta_\mathrm{m}$, with $\beta=(k_\mathrm{B}T)^{-1}$,
we find
\begin{equation}
\frac{\partial A_\mathrm{eff}}{\partial\ln c_\mathrm{L}}
=
\frac{1}{2}
\frac{\partial^2(\rho\theta_\mathrm{m})}{\partial\rho^2}
=
\frac{\theta_\mathrm{m}}{2\rho}
\left[
\alpha_\mathrm{m}(1+\alpha_\mathrm{m})
+
\frac{\partial\alpha_\mathrm{m}}{\partial\ln\rho}
\right].
\label{eq:SelectivityAttraction}
\end{equation}
Integrating this relation over linker concentration, with \(A_{\mathrm{eff}}\to0\) as \(c_{\mathrm L}\to0\), gives the effective attraction from adsorption isotherms at different site densities. This relationship follows from thermodynamic identities alone and is independent of the molecular mean-field approximations introduced above.

\begin{figure}[ht!]
\includegraphics[trim=0 0cm 0 0, width=8.5cm]{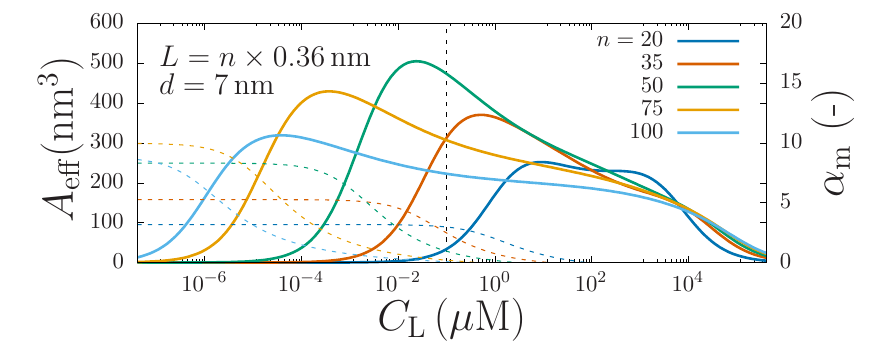}
\includegraphics[width=8.5cm]{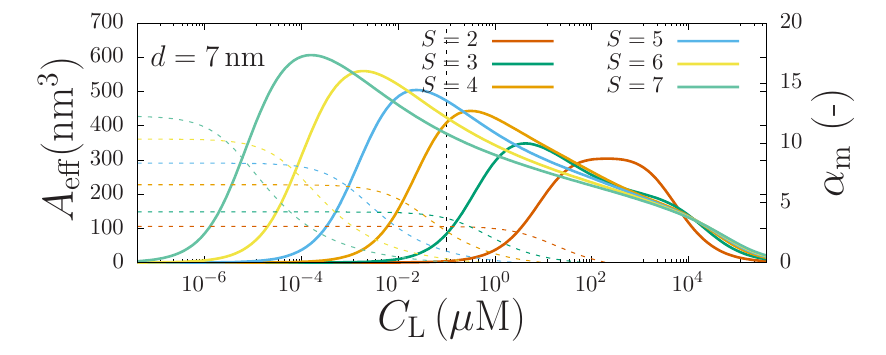}
\caption{Effective attraction (solid curves) and adsorption
selectivity (dashed curves) as functions of free-linker
concentration for the EPYC1--Rubisco system.
The vertical line indicates the estimated physiological
free-linker concentration.
The top and bottom panels show the influence of spacer
length and sticker number, respectively.
Spacer length is expressed in the number of amino acids
$n=L/(0.36\,\mathrm{nm})$,
where $0.36\,\mathrm{nm}$ is the contour length per residue
\cite{payne-dwyer_predicting_2024}.}\label{fig:Interactions}
\end{figure}

The relationship between the effective attractions  (solid curves) and selectivity factor (dashed curves) is illustrated by considering the EPYC1--Rubisco system in Fig.~\ref{fig:Interactions}.
In the dilute-linker limit, $A_\mathrm{eff}$ increases linearly with $c_\mathrm{L}$, with a slope determined by the density dependence of adsorption. As the linker concentration increases, the attraction can continue to grow even as competition for binding sites suppresses superselectivity. Appreciable attractions therefore persist into the regime of competitive binding. At sufficiently high linker concentrations, however, monovalent binding dominates \cite{payne-dwyer_predicting_2024}, and the grand potential $\omega/k_\mathrm{B}T\simeq-\ln(Sc_\mathrm{L})$ becomes independent of site density, such that $A_\mathrm{eff}\rightarrow0$. Thus, superselectivity contributes to the emergence of linker-mediated attractions without directly determining their magnitude.

The strength and concentration range of these attractions depend on the molecular architecture of the linkers. At the estimated wild-type free-linker concentration of approximately $0.1\,\mathrm{\mu M}$, the continuum theory predicts that the native spacer length is close to optimal for linker-mediated attraction (top panel). At this concentration, appreciable attraction emerges for linkers with at least four stickers, consistent with recent experimental observations of the minimum valency required for pyrenoid formation in vivo \cite{Kumar2026.01.27.701992}. Increasing the sticker number further has relatively little effect on the interaction strength at this concentration (bottom panel).

Incorporating the spatial statistics and interactions of mobile Rubisco molecules into this framework provides a route from molecular binding statistics to predictions of phase coexistence.

\paragraph{Conclusion --}~
We have developed a mean-field theory of multivalent polymer binding by coarse-graining our previously developed molecular model \cite{payne-dwyer_predicting_2024}, and validated it against Monte Carlo simulations. The theory predicts crossovers between monovalent, superselective and competitive binding, for which we have derived scaling relationships. We have also established a general thermodynamic relationship between adsorption selectivity and linker-mediated attraction. This shows that appreciable attractions can continue to increase as competition for binding sites suppresses superselectivity, before eventually disappearing at high linker concentrations.

Applied to the EPYC1--Rubisco pyrenoid, the theory predicts that the native spacer length is close to optimal for linker-mediated attraction at the estimated physiological concentration. Increasing sticker number strongly affects the concentration range for these attractions, while the interaction strength becomes less sensitive to sticker number at higher concentrations. These results show how the molecular architecture of multivalent proteins can influence the interactions that favour condensation, and provide a starting point for incorporating molecular binding statistics into more complete descriptions of phase separation.

\section{Acknowledgements}~
CS is grateful for  fruitful discussions on superselectivity with 
Mark Miller (Durham) and Ralf Richter (Leeds), and on the properties of pyrenoids  with Mark Leake (Durham) and Luke Mackinder (York). 
CS was funded by the EPSRC [Grant no. EPSRC
(EP/W024063/1)].

\section{Data Availability}
All simulation data and reproduction scripts will be made available on Zenodo. 
The simulation software itself is available on request.

\bibliography{sorsamp}%

%
%
\newpage
\onecolumngrid

\setcounter{equation}{0}
\renewcommand{\theequation}{S\arabic{equation}}
\setcounter{page}{1}
\renewcommand{\thepage}{S\arabic{page}}
\setcounter{figure}{0}
\renewcommand{\thefigure}{S\arabic{figure}}

\vspace{2em}
\begin{center}
\textbf{SUPPLEMENTARY INFORMATION}
\end{center}

\vspace{2em}
\noindent This Supplementary Information provides the details by which the molecular partition sum is coarse-grained, and how the thermodynamic limit and saddle-point approximation are applied. Finally, it will provide further numerical verification of the theory against our simulations; in particular through the measures of the grand potential per site and the effective attractions.

\vspace{2em}

\noindent \textbf{Coarse-graining the molecular partition sum}

Our first approximation is the factorisation of the
conformational weights of the bound molecules,
\begin{equation}
    W_{\mathbf m}\simeq\prod_{b=1}^{S}w_b^{m_b}.
\end{equation}
This approximation neglects correlations between the
conformations of different molecules, while the restriction
that only one sticker can bind per site is retained through
the combinatorial factor $N!/(N-B)!$ in the partition sum.

Our second approximation is that the conformations of the
individual constrained spacers of a molecule can be treated
independently. If stickers $i_1<i_2<\cdots<i_b$ are bound,
the elastic free energy of the molecule is given by the sum
of the free energies of the $b-1$ constrained spacers.
We therefore approximate the conformational weight by
\begin{equation}
    w_b\simeq
    \frac{1}{\binom{S}{b}}
    \sum_{i_1<\cdots<i_b}
    \prod_{j=1}^{b-1}
    \left\langle
    \exp\left(
        -\frac{G(z;L_{i_j,i_{j+1}})}{k_\mathrm{B}T}
    \right)
    \right\rangle_z.
    \label{eq:SI_wb}
\end{equation}
This approximation neglects spatial correlations arising
from the restriction that nonconsecutive bound stickers
of the same molecule cannot occupy the same site.

Our third approximation is that the binding sites are
distributed uniformly throughout a volume $V$, with number
density $\rho=N/V$ and characteristic intersite distance
$d=\rho^{-1/3}$. To account approximately for the finite
separation between binding sites, we exclude spacer
extensions smaller than $d$, such that
\begin{equation}
    \left\langle
    \exp\left(-\frac{G(z;L)}{k_\mathrm{B}T}\right)
    \right\rangle_z
    \simeq\frac{v_L}{V},
\end{equation}
where we have introduced the conformational volume
\begin{equation}
    v_L=4\pi\int_d^L\mathrm{d}z\,z^2
    \exp\left(-\frac{G(z;L)}{k_\mathrm{B}T}\right),
    \label{eq:SI_ConformationalVolume}
\end{equation}
which vanishes if $d\geq L$. In the main text, we have shown this integral leads to accurate predictions for the molecular coverages. Later in this Supplementary Information show these also lead to reasonable predictions for the effective interactions, whose sensitivity to the continuum approximation is investigated below.

As a consequence of these approximations, the conformational
weight of a molecule scales as $w_b\propto N^{1-b}$ at fixed
site density. It is therefore convenient to introduce the
intensive statistical weight
\begin{align}
    q_b&\equiv N^{b-1}\binom{S}{b}w_b
    \notag\\
    &\simeq
    \sum_{i_1<\cdots<i_b}
    \prod_{j=1}^{b-1}\rho v_{L_{i_j,i_{j+1}}}.
    \label{eq:SI_qb}
\end{align}
Note that the contour length of a constrained spacer depends
on the number of unbound stickers that it contains.
Moreover, the contour lengths of successive constrained
spacers are correlated due to the finite contour length
of the molecule.

To evaluate Eq.~(\ref{eq:SI_qb}) efficiently, we introduce
the dimensionless statistical weight associated with a
constrained spacer connecting stickers $i$ and $j$,
\begin{equation}
    Q_{ij}=\rho v_{L_{i,j}}.
\end{equation}
We further introduce the quantity $T_{i,b}$, which is the
total statistical weight of configurations in which a
molecule is bound through $b$ stickers, the last of which
is sticker $i$ along the contour of the chain.
For singly bound molecules we have $T_{i,1}=1$, while for
$b>1$ we obtain the recursion relation
\begin{equation}
    T_{i,b}=\sum_{j=1}^{i-1}T_{j,b-1}Q_{ji},
    \label{eq:SI_SpacerRecursion}
\end{equation}
where $T_{i,b}=0$ if $i<b$.
Summation over all possible choices of the last bound
sticker yields
\begin{equation}
    q_b=\sum_{i=b}^{S}T_{i,b}.
    \label{eq:SI_qbRecursion}
\end{equation}
Importantly, this recursion introduces no additional
physical approximations, but reduces the computational
complexity of evaluating all conformational weights from
exponential to cubic in the number of stickers.
For molecules with regularly spaced stickers,
$L_{i,j}=(j-i)L$, only $S-1$ distinct conformational
volumes need to be evaluated.

Finally, using the identities $\sum_b m_b=M$ and
$\sum_b bm_b=B$, we obtain the coarse-grained partition sum
\begin{equation}
    Z_{M,B}\simeq
    \frac{N!}{(N-B)!N^{B-M}}
    \sum_{\mathbf m\in\mathcal P_{M,B}}
    \prod_{b=1}^{S}\frac{q_b^{m_b}}{m_b!}.
    \label{eq:SI_ZMBapprox}
\end{equation}

\vspace{2em}

\noindent \textbf{Thermodynamic limit and saddle-point approximation}

To evaluate the partition sum in the thermodynamic limit,
we introduce the number of bound molecules and stickers
per site, $\theta_\mathrm{m}=M/N$ and
$\theta_\mathrm{s}=B/N$, respectively, and the number of
molecules per site that are bound using $b$ stickers,
$x_b=m_b/N$.
These quantities obey
\begin{equation}
    \theta_\mathrm{m}=\sum_{b=1}^{S}x_b,
    \qquad
    \theta_\mathrm{s}=\sum_{b=1}^{S}bx_b.
    \label{eq:SI_Constraints}
\end{equation}

Denoting the contribution of an individual integer
partition to Eq.~(\ref{eq:SI_ZMBapprox}) by $Z_{\mathbf m}$,
we have
\begin{align}
    \ln Z_{\mathbf m}={}&
    \ln N!-\ln(N-B)!-(B-M)\ln N
    \notag\\
    &+\sum_{b=1}^{S}
    \left[m_b\ln q_b-\ln m_b!\right].
\end{align}
Our fourth approximation is to invoke Stirling's
approximation, $\ln n!\simeq n\ln n-n$, which yields
\begin{align}
    \frac{\ln Z_{\mathbf m}}{N}
    \simeq{}&
    -(1-\theta_\mathrm{s})\ln(1-\theta_\mathrm{s})
    -\theta_\mathrm{s}+\theta_\mathrm{m}
    \notag\\
    &-\sum_{b=1}^{S}x_b
    \ln\left(\frac{x_b}{q_b}\right).
    \label{eq:SI_Stirling}
\end{align}
Note that all terms proportional to $\ln N$ cancel,
such that the logarithm of the contribution of an
individual integer partition is extensive.

Finally, we invoke a saddle-point approximation.
At fixed $S$, the number of integer partitions in
$\mathcal P_{M,B}$ grows at most polynomially with $N$,
whereas their statistical weights scale exponentially
with $N$. Indeed,
\begin{equation}
    \max_{\mathbf m}Z_{\mathbf m}
    \leq Z_{M,B}
    \leq |\mathcal P_{M,B}|
    \max_{\mathbf m}Z_{\mathbf m},
\end{equation}
within the coarse-grained description.
Since $|\mathcal P_{M,B}|\leq(N+1)^S$, the contribution
of the number of integer partitions to $\ln Z_{M,B}/N$
vanishes in the thermodynamic limit. Consequently,
\begin{equation}
    \lim_{N\rightarrow\infty}\frac{\ln Z_{M,B}}{N}
    =
    \lim_{N\rightarrow\infty}
    \max_{\mathbf m\in\mathcal P_{M,B}}
    \frac{\ln Z_{\mathbf m}}{N},
    \label{eq:SI_SaddlePoint}
\end{equation}
where the thermodynamic limit is taken at fixed site
density $\rho$ and coverages $\theta_\mathrm{m}$
and $\theta_\mathrm{s}$.

Including the binding free energy $\varepsilon$ per
bound sticker, we introduce the free energy per site
associated with a distribution $\mathbf x=[x_1,\ldots,x_S]$,
\begin{align}
    \frac{f(\mathbf x)}{k_\mathrm{B}T}={}&
    (1-\theta_\mathrm{s})\ln(1-\theta_\mathrm{s})
    +\theta_\mathrm{s}-\theta_\mathrm{m}
    \notag\\
    &+\sum_{b=1}^{S}x_b
    \ln\left(\frac{x_b}{q_b}\right)
    +\frac{\varepsilon}{k_\mathrm{B}T}\theta_\mathrm{s}.
    \label{eq:SI_FreeEnergy}
\end{align}
The equilibrium free energy per binding site is then
given by
\begin{equation}
    \frac{F}{N}=\min_{\mathbf x}f(\mathbf x),
\end{equation}
where the minimisation is performed subject to the
constraints in Eq.~(\ref{eq:SI_Constraints}), with
$x_b\geq0$.

Thus, the summation over integer partitions in the
microscopic description is replaced by a variational
problem in the thermodynamic limit.
For a system exchanging linker molecules with a reservoir,
the constraints on $\theta_\mathrm{m}$ and
$\theta_\mathrm{s}$ are relaxed, and equilibrium is
obtained by minimising the grand potential per site,
$w(\mathbf x)=f(\mathbf x)-\mu\theta_\mathrm{m}$,
subject to $\theta_\mathrm{s}\leq1$ and $x_b\geq0$,
as described in the main text.

\vspace{2em}
\noindent \textbf{High-density crossover regimes}

To obtain the crossover concentrations in the
high-site-density limit, we introduce
$x=K(1-\theta_\mathrm{s})$, such that
Eqs.~(16) and (17) of the main text become
\begin{align}
    K &= x+Sc_\mathrm{L}x(1+x)^{S-1},\\
    \theta_\mathrm{m}
    &=\frac{c_\mathrm{L}}{K}
    \left[(1+x)^S-1\right].
\end{align}
Since $K\propto\rho$ in the high-density limit,
the selectivity factor is obtained by logarithmic
differentiation with respect to $K$ at fixed
$c_\mathrm{L}$, yielding
\begin{equation}
    \alpha_\mathrm{m}
    =-1+
    \frac{Sx}{1+Sx-(S-1)x^2/K}
    \frac{1}{1-(1+x)^{-S}}.
\end{equation}
For $x\gg1$, this reduces to
\begin{equation}
    \alpha_\mathrm{m}\simeq
    \frac{(S-1)x^2/K-1}
    {1+Sx-(S-1)x^2/K}.
\end{equation}
The corresponding linker concentration is
\begin{equation}
    c_\mathrm{L}\simeq\frac{K-x}{Sx^S}.
\end{equation}

At the onset of competitive adsorption
($\alpha_\mathrm{m}=0$), we obtain
$x\simeq\sqrt{K/(S-1)}$, such that
\begin{equation}
    c_\mathrm{L}^*\simeq
    \frac{(S-1)^{S/2}}{SK^{(S-2)/2}},
\end{equation}
which recovers Eq.~(18) of the main text.

At the crossover from superselective to
sublinear adsorption ($\alpha_\mathrm{m}=1$),
we instead obtain, for $S>2$,
$x\simeq SK/[2(S-1)]$, yielding
\begin{equation}
    c_\mathrm{L}^*\simeq
    \frac{(S-2)[2(S-1)]^{S-1}}
    {S^{S+1}K^{S-1}},
\end{equation}
which recovers Eq.~(19) of the main text.

\vspace{2em}
\noindent \textbf{Numerical validation}

In the main text, we have verified that the mean-field theory 
accurately describes the adsorption isotherms.
Here, we show that the underlying grand potential is indeed 
described with similar accuracy, while the effective interactions
(which are derived from the density curvature of this potential)
are more sensitive to the local site distributions.
Thus, in the following, we use both the continuum integral
description of $v_L$ and a discrete summation corresponding
to the regular cubic lattice used in our simulations,
\begin{equation}
    \rho v_L \rightarrow
    \sum_{\mathbf n\in\mathbb Z^3\setminus\{\mathbf 0\}}
    \exp\left[
        -\frac{G(d|\mathbf n|;L)}{k_\mathrm{B}T}
    \right],
\end{equation}
where the summation is restricted to sites within the
contour length of the spacer.

\begin{figure}[ht!]
    \includegraphics[width=8 cm]{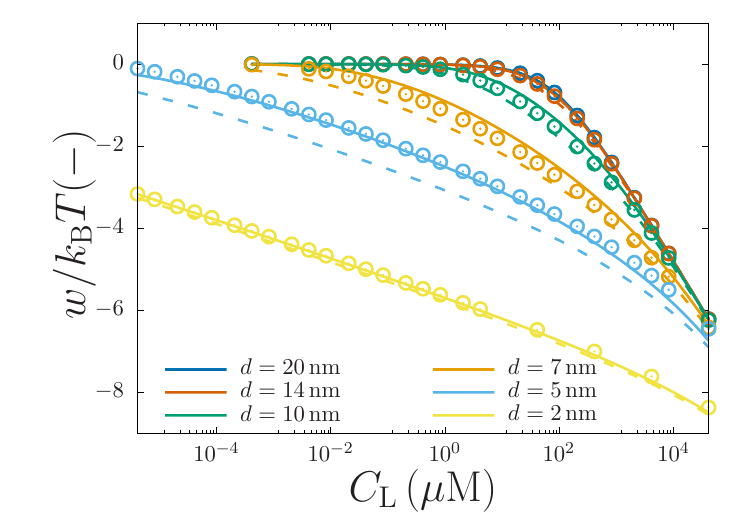}
    \includegraphics[width=8 cm]{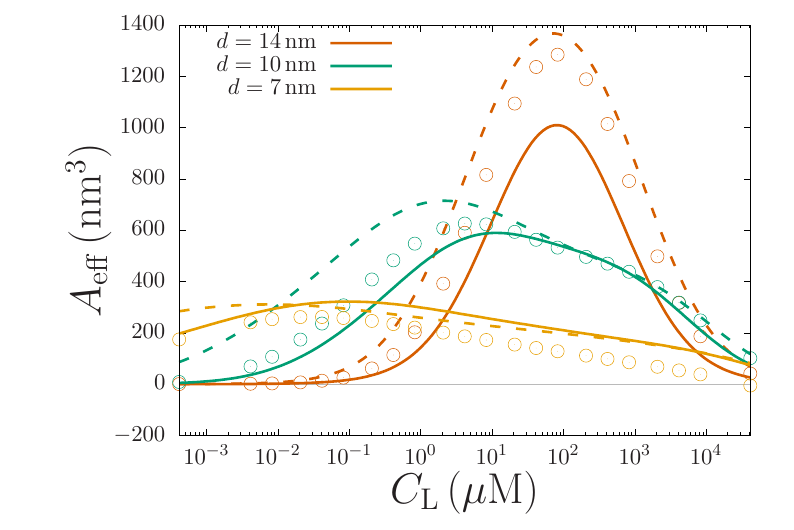}
    \caption{Numerical validation of the calculated grand 
    potential (left) and effective attraction (right).
    Symbols represent simulation results, while solid
    and dashed curves represent the mean-field theory
    using the continuum integral and discrete lattice
    summation, respectively.
    The effective attraction is estimated using identical
    three-point density intervals for the simulations
    and theory.}
    \label{fig:SI_GrandPotential}
\end{figure}

From our simulations, we calculate the grand potential
using thermodynamic integration,
\begin{equation}
    \frac{\omega(\rho,c_\mathrm{L})}{k_\mathrm{B}T}
    =
    -\int_0^{c_\mathrm{L}}
    \theta_\mathrm{m}(\rho,c')
    \,\mathrm{d}\ln c'.
\end{equation}
The results are compared with the mean-field predictions
in the left panel of Fig.~\ref{fig:SI_GrandPotential}.
For $d=2\,\mathrm{nm}$, the lowest simulated linker
concentrations already correspond to near-saturation
by fully multivalently bound linkers. We therefore
estimate the missing low-concentration contribution
to the thermodynamic integral using the mean-field
theory restricted to fully bound molecules, yielding
a constant shift of the simulated grand potential.

The effective attraction (right panel) is subsequently
estimated from the grand potential at three successive
site densities, using a finite-difference approximation of
\begin{equation}
    A_\mathrm{eff}=
    -\frac{1}{2}
    \left.
    \frac{\partial^2}{\partial\rho^2}
    \left(\frac{\rho\omega}{k_\mathrm{B}T}\right)
    \right|_{c_\mathrm{L}}.
\end{equation}
For a direct comparison, we apply the same three-point
finite-difference scheme to the simulated and theoretical
grand potentials. The relatively large density intervals
affect both estimates and explain the differences from
the local free-energy curvatures presented in the main text.

Interestingly, replacing the continuum integral by a
discrete lattice summation does not systematically
improve agreement of the grand potential with the simulations.
This may reflect a partial compensation between the
continuum approximation and the neglect of spatial
correlations between constrained spacers.
While the two descriptions yield similar adsorption
thermodynamics, quantitative differences are more
pronounced for the effective attraction, which is
determined by the density curvature of the grand
potential.
Nevertheless, both descriptions capture the characteristic
concentration dependence of the effective attraction
observed in the simulations.
For the application to biomolecular condensates,
where the binding sites are not arranged on a
regular lattice, we retain the continuum description.

\end{document}